\documentclass[preprint]{aastex631}

\usepackage[utf8]{inputenc}
\usepackage{CJK}
\usepackage{amsmath}

\usepackage{booktabs}
\usepackage{tablefootnote}

\newcommand{\ghz}{\ensuremath{\,{\rm GHz}}}
\newcommand{\mum}{\ensuremath{\,{\rm \mu m}}}
\newcommand{\pc}{\ensuremath{\,{\rm pc}}}
\newcommand{\kpc}{\ensuremath{\,{\rm kpc}}}

\newcommand{\K}{\ensuremath{\,{\rm K}}}

\newcommand{\s}{\ensuremath{\,{\rm s}}}
\newcommand{\g}{\ensuremath{\,{\rm g}}}
\newcommand{\mm}{\ensuremath{\,{\rm mm}}}

\newcommand{\jy}{\ensuremath{\,{\rm Jy}}}
\newcommand{\msun}{\ensuremath{\,{M_\odot}}}
\newcommand{\kms}{\ensuremath{\,{\rm km\, s^{-1}}}}
\newcommand{\erg}{\ensuremath{\,{\rm erg}}}

\newcommand{\ic}{IC\thinspace 443}

\newcommand{\AV}{\ensuremath{{A_V}}}
\newcommand{\RV}{\ensuremath{R_V}}
\newcommand{\CO}{\ensuremath{\rm ^{12}CO}}
\newcommand{\EBV}{\ensuremath{E(B-V)}}
\newcommand{\Kextv}{\ensuremath{K_V^{\rm ext}}}

\newcommand{\tw}{\ensuremath{T_{\rm w}}}
\newcommand{\mw}{\ensuremath{M_{\rm w}}}
\newcommand{\tc}{\ensuremath{T_{\rm c}}}
\newcommand{\mc}{\ensuremath{M_{\rm c}}}

\newcommand{\spitzer}{\emph{Spitzer}}
\newcommand{\wise}{\emph{WISE}}
\newcommand{\iras}{\emph{IRAS}}
\newcommand{\akari}{\emph{AKARI}}
\newcommand{\planck}{\emph{Planck}}

\defcitealias{gal14}{G14}
\defcitealias{car89}{CCM89}
\defcitealias{fit99}{F99}

\usepackage{multirow}

\shorttitle{Dust in SNR IC\thinspace 443}
\shortauthors{Li et al.}

\graphicspath{{./}{figures/}}

\begin{document}
\begin{CJK*}{UTF8}{gbsn}


\title{Dust Mass Associated with the Supernova Remnant IC\thinspace 443 when Emission Meets Extinction}

\author[0000-0001-9328-4302]{Jun Li (李军)}
\affiliation{Department of Astronomy, Beijing Normal University, Beijing 100875, Peopleʼs Republic of China}
\email{lijun@mail.bnu.edu.cn}

\correspondingauthor{Biwei Jiang}
\email{bjiang@bnu.edu.cn}

\author[0000-0003-3168-2617]{Biwei Jiang (姜碧沩)}
\affiliation{Department of Astronomy, Beijing Normal University, Beijing 100875, Peopleʼs Republic of China}

\author[0000-0003-2645-6869]{He Zhao (赵赫)}
\affiliation{Department of Astronomy, Beijing Normal University, Beijing 100875, Peopleʼs Republic of China}
\affiliation{University C\^ote d'Azur, Observatory of the C\^ote d'Azur, CNRS, Lagrange
Laboratory, Observatory Bd, CS 34229, \\
06304 Nice cedex 4, France}



\begin{abstract}

The dust mass of the well-known supernova remnant (SNR) \ic\ is estimated from both the infrared emission and the visual extinction. With photometry to the images taken by \spitzer, \wise, \iras, \akari\ and \planck, the spectral energy distribution (SED) of the dust is obtained after subtracting the synchrotron radiation and considering the spectral line emission. The dust mass is derived from fitting the SED by a two-component model, which results in a warm component of the temperature of $\sim$53\K\ and the mass of 0.1\msun\, and a cold component of the temperature of $\sim 17$\K\ and the mass of 46\msun. On the other hand, the dust mass is derived to be $\sim$ 66\msun\ from the visual extinction of \ic\ which is identified from the 3D Bayestar extinction map and its coincidence with the infrared emission morphology. Roughly the dust mass derived from the infrared emission and the extinction agree mutually. However, the dust mass derived from the infrared emission can be adjusted to be more consistent with that from the extinction by using different dust opacity property or considering optically thick radiation. In addition, the distribution of dust temperature and mass is analyzed by fitting the SED pixel by pixel.

\end{abstract}



\section{Introduction}
\label{sec:intro}

Supernova (SN) explosion is a violent process with an energy release on the order of 10$^{51}$\erg. On one hand, dust can form in the ejected matter when SN cools down. Theoretically, a core-collapse SN would contribute 0.1--1\msun\ of dust grains to the interstellar space \citep{Nozawa2003}. The measurements of the very young remnant SN\thinspace 1987A infer the mass of dust formed in the SN ejecta to be around a few tenths solar mass \citep{Matsuura2015}. On the other hand, the shocks and large amount of ejecta produced by a SN also interact violently with the interstellar material \citep{Burton1988,Rho2001,Ritchey2020}. Dust in the ambient surrounding is destroyed by the strong SN shocks. \cite{Lakicevic2015} presented a systematic study of the complete sample of SNRs in the LMC and derived that a supernova would remove 3.7\msun\ of dust on average.

No matter whether SN as a contributor or a destroyer to interstellar dust, the quantitative determination of the SNR dust is crucial to understanding the cycle of cosmic dust, yet it suffers some uncertainties. The popular method to estimate the dust mass of SNRs is to fit the observational spectral energy distribution (SED) by some dust models \citep{Hildebrand1983,Gomez2012,Matsuura2015}, in which a few factors can bring about uncertainties. Because there is no information on the distance of the emitters, the contribution from the foreground and/or background to the infrared radiation can hardly be separated from the radiation of the SNR itself, which would lead to over-estimating the dust mass. Besides, the dust mass is very sensitive to the temperature that depends on the wavelength range of the SED, in particular the far-infrared (FIR) or sub-millimeter (submm) band as the key waveband is generally missing \citep{Matsuura2015}. In addition, the temperature also depends on the dust model. Another technique that has been used to estimate the dust mass uses the SNR-produced extinction since the extinction is proportional to the dust column density \citep{Zhao2018,Wang2020}. The extinction method has the advantage of building a three-dimensional extinction map so that the foreground or background clouds can be recognized \citep{Zhao2020,Green2019}, and it neither depends on the dust temperature. However, the extinction method cannot distinguish the dust of the SNR  from the surrounding molecular cloud that is very possibly associated with the SNR.

In this work, we try to find a consistent result of the SNR dust mass calculated from the emission and the extinction respectively. For this purpose, \ic\ (G189.1+3.0) is chosen. \ic\ is one of the most studied SNRs. Its distance was early determined to be 1.5\kpc\ from its progenitor likely being a member of the Gem OB1 association \citep{Braun1986,Humphreys1978}. However, the new measurements lead to another concensus. \citet{Ritchey2020} found the distance to the Gem OB1 association  in the range of 1.8-2.3\kpc\ with the Gaia/DR2 astrometric information. \cite{Ambrocio-Cruz2017} determined the distance of \ic\  to be 1.9\kpc\ from the kinematic study of the H$\alpha$ line. \citet{Yu2019} and \citet{Zhao2020} also estimated its distance to be 1.7-1.8\kpc\ based on the dust extinction jump along the sightline. Thus, a distance of 1.8\kpc\ is adopted in this work. The angular diameter of $\sim 45'$ \citep{Green2019snr} at 1.8\kpc\ then converts to a linear one of 23.56\pc, i.e. 0.52\pc\ per arcmin. This size is consistent with the argument that \ic\ is an evolved SNR with an age of 3--30\,kyr \citep{Petre1988,Chevalier1999,Olbert2001}. \cite{Lee2008} estimated the age to be $\sim20$\,kyr by assuming \ic\ as a breakout SNR. Recently, \cite{Ustamujic2021} developed a detailed 3D hydrodynamic simulation to model the evolution of \ic, and derived an age of 8\,kyr.

The morphology and evolution of SNRs depend on their surrounding environment. A lot of studies show that \ic\ has been evolving in a non-uniform-distributed medium \citep[e.g.,][]{Mufson1986,Braun1986}. In both optical and radio regimes, \ic\ appears to have similar morphology that is consisted of two hemispheres with different radius and centroid \citep[namely shell A and B by the nomenclature from][]{Braun1986}. Shell A locates in the northeastern (NE) region, forms a nearly circular rim and dominates the IR structure. In the southwestern (SW) region (Shell B), \ic\ is larger and more extended, which is suggested to be a breakout part of the remnant into the low-density medium \citep{Lee2008}. In the infrared regime, the morphology of \ic\ is not well correlated with that in the radio/optical \citep{Mufson1986}, implying different and complex physical conditions in its spatial distribution.
The X-ray emission of \ic\ is not shell-like, but centrally peaked and filling the entire SNR according to the XMM-Newton images \citep{Troja2006,Troja2008}. The simulations by \cite{Ustamujic2021} reproduced the complex X-ray morphology of \ic\ and clearly illustrated its evolution in a highly inhomogeneous medium.

The study of dust in \ic\ has been conducted before. \cite{Mufson1986} found the shock-heated dust with a temperature of $\sim$42--45\K\ in \ic\ by fitting the \iras\ photometric data. \cite{Kokusho2013} reproduced its emission measured by \emph{Spitzer} and \akari\ with a two-temperature modified blackbody model, and found the dust temperature of 56.3\K\ and 14.8\K\ respectively associated with \ic. The warm component agrees with the result of  \citet{Mufson1986}, while the cold one has a too low temperature even when compared to the general ISM and may not associate with \ic. In addition to heating grains to hotter temperatures,  the SNR shocks can also destroy small grains efficiently in the post-shock gas and modify the pre-shock grain size distribution. Early observation of strong [\mbox{Ca\,{\sc ii}}] lines in \ic\ may suggest that the interstellar grains are partially destroyed by the shocks \citep{Fesen1980}.
\cite{Kokusho2013,Kokusho2015} found an anti-correlation between the [\mbox{Fe\,{\sc ii}}] and dust emission, indicating interstellar dust destruction by the SNR shocks. \cite{Ritchey2020} also reported enhanced gas-phase Ni and Ca abundances along one line-of-sight through the NE shell, which indicates significant dust destruction via grain-grain collisions or thermal sputtering.

The paper is organized as follows. In Section \ref{sec:emission} we analyze the mid-infrared (MIR) to submm emission, in particular regarding the dust temperature and mass. In Section \ref{sec:extinction} we present the study the dust mass of \ic\ based on the extinction map. Section \ref{sec:unification} compares the dust mass derived from emission with extinction. Section \ref{sec:conclusion} summarizes the results.

\section{Dust Mass Derived from the Infrared Emission} \label{sec:emission}

\subsection{Archival Data and Reduction} \label{sec:data}

\subsubsection{The Data}
In order to get a broad wavelength coverage of the SED, we collect the imaging observations of \ic\ from mid-infrared (MIR) to millimeter (mm) wavebands. Specifically, the following images are used:
\begin{itemize}
  \item The archival PBCD data of the \spitzer\ observations with Multiband Imaging Photometer (MIPS) \citep{Rieke2004} at 24, 70 and 160\mum\ observed in October 2004 (PI: George Rieke) from the \spitzer\ Heritage Archive\footnote{https://sha.ipac.caltech.edu/applications/Spitzer/SHA/}.
  \item The \emph{Wide field Infrared Survey Explorer} (\wise) images in four bands at 3.4, 4.6, 12 and 22\mum\ \citep{Wright2010}, in which only the later two bands are used in our analysis.
  \item The \emph{Infrared Astronomical Satellite} (\iras) all-sky survey images at 12, 25, 60 and 100\mum. The calibration uncertainty for all the four \iras\ bands is set to $20\%$.
  \item The \akari/Far-infrared All-Sky Survey Maps at 65\mum\ (N60), 90\mum\ (WIDE-S), 140\mum\ (WIDE-L) and 160\mum\ (N160) \citep{Doi2015}.
  \item The \planck\ all-sky survey maps in all the nine wavebands \citep{Planck2016}.  The 4 high-frequency (i.e. 350, 550, 850 and 1380\mum) images are used.
\end{itemize}

\begin{figure*}
	\centering
	\includegraphics[scale=0.7]{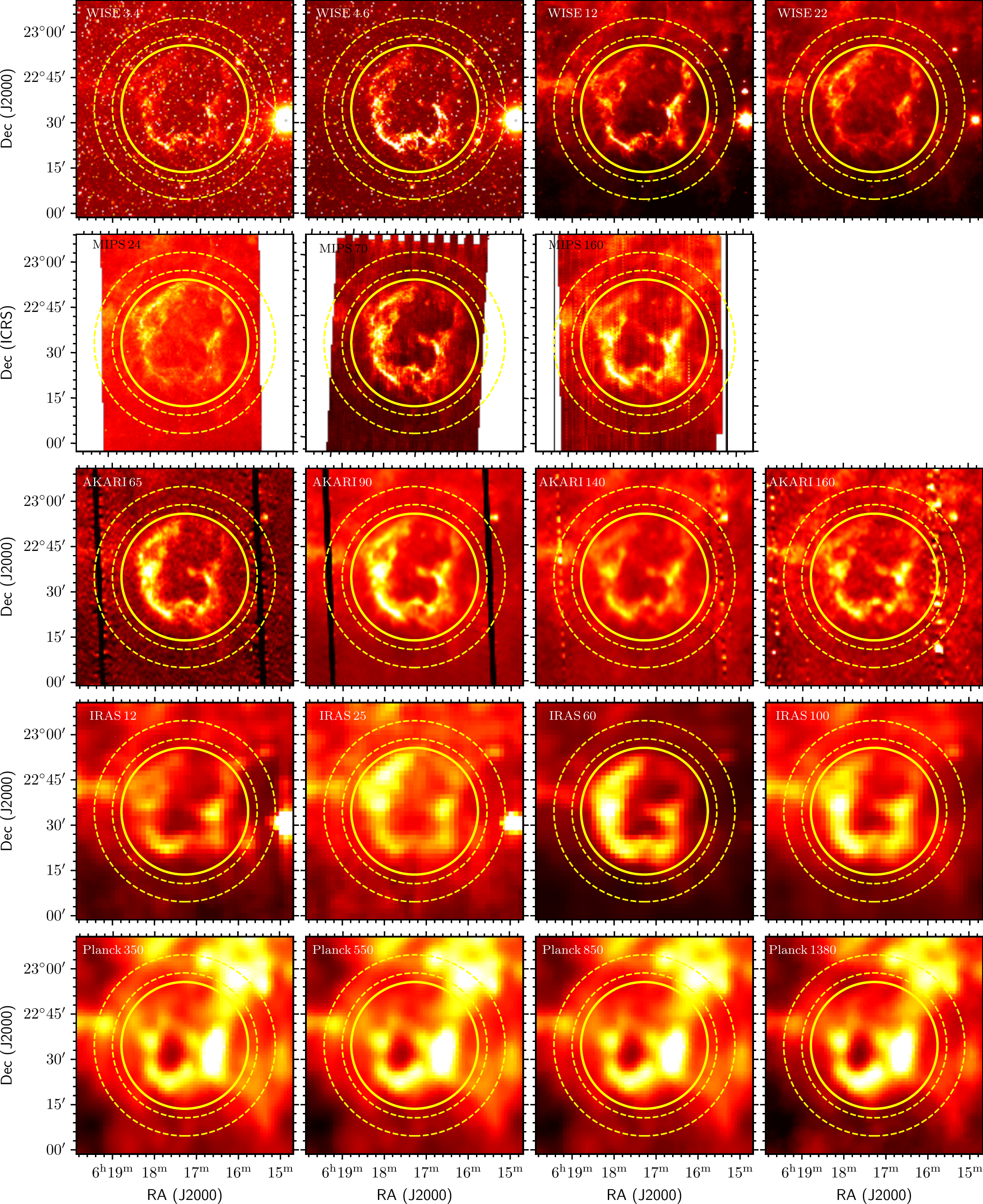}
	\caption{The multi-band images of \ic\ centered at $\rm R.A.=6h17m16.635s, Dec.=+22^\circ34'43.0''$ in the \wise\ 3.4, 4.6, 12 and 22\mum\ bands (first row), \spitzer/MIPS 24, 70 and 160\mum\ bands (second row), \akari\ 65, 90, 140 and 160\mum\ bands (third row), \iras\ 12, 25, 60 and 100\mum\ bands (fourth row), and \planck\ 350, 550, 850 and 1380\mum\ bands (fifth row). Each image is plotted in its original resolution and pixel size. The aperture for the integrated photometry is denoted by the yellow solid circle with a radius of 21\arcmin\ and  the yellow dashed lines denote the sky background annulus of the inner and outer radii of 24\arcmin\ and 30\arcmin.
	\label{fig:all_images} }
\end{figure*}

In summary, we take the images of \spitzer/MIPS 24, 70, 160\mum, \wise\ 12, 22\mum, \iras\ 12, 25, 60, 100\mum, \akari\ 60, 90, 140, 160\mum\ and \planck\ 350, 550, 850, 1380\mum\ spanning the range from $\sim10\mum$ to 1.4\mm. The \wise, \iras, \akari\ and \planck\ images are retrieved from NASA/IPAC Infrared Science Archive\footnote{https://irsa.ipac.caltech.edu/frontpage/}. These images are displayed in Figure \ref{fig:all_images}.

\subsubsection{Infrared Photometry}

The fluxes of \ic\ are measured via the aperture photometry by the Photutils\footnote{https://photutils.readthedocs.io/} Python package.  The adopted apertures of photometry for the SNR and the background are the same for all images and displayed in Figure \ref{fig:all_images} by solid and dashed circles respectively. The integrated fluxes are measured using a circular aperture with a radius of 21\arcmin, centered at $\rm R.A.=6h17m16.635s, Dec.=+22^\circ34'43.06"$. The background emission is estimated by taking the median value within the background annulus whose inner and outer radius is 24\arcmin\ and 30\arcmin, respectively. The median value is hardly influenced by the bright emission in the annulus in comparison with the average background.
We do not apply any color corrections or aperture corrections, because the aperture corrections are small at such large aperture size. The color corrections are also small, being no more than a few percent in all cases. The details and the results with the uncertainty of photometry are summarized in Table \ref{tab:data} for the wavebands we used. The flux errors are dominated by the calibration uncertainty for the wavebands with $\lambda\leq 160\mum$. The flux errors of the \planck\ bands are dominated by both calibration uncertainty and sky background errors which is evaluated by the standard deviation of the fluxes from the background annulus.

Our results are compared with previously published fluxes as listed in Table \ref{tab:data}, which involves the wavebands of the \spitzer/MIPS 24 and 70\mum\ \citep{Koo2016}, \iras\ 12, 25, 60 and 100\mum\ \citep{Mufson1986}, and \planck\ 350, 550, 850, 1380\mum\ \citep{Planck2016}.  The fluxes of MIPS and \iras\ measured in this work are consistent with previous works. While the \planck\ fluxes are visibly smaller than that reported by \cite{Planck2016}. This discrepancy can be understood by that they adopted a larger aperture for photometry.\footnote{If we take the aperture radius $\theta_{\rm ap}$ of 30.2\arcmin\, and the sky background in an annulus with the inner and outer radii of $1.5\,\theta{\rm ap}$ and $2\,\theta{\rm ap}$ as in \cite{Planck2016}, the derived flux would become $2345\pm159$, $878\pm56$, $256\pm8$ and $63\pm2$\jy\ for \planck\ 350, 550, 850 and 1380\mum\ bands, which in general agree with the measurements within the uncertainties by \cite{Planck2016}, though with about about 8\% higher flux.} Within our measurements, the integrated fluxes in similar bands coincide with each other very well,  e.g. the MIPS 24\mum\ flux is in good agreement with \wise\ 22\mum\ and \iras\ 25\mum\ flux.

\subsection{Analysis of the Dust Emission}

\subsubsection{The Multi-wavelength Morphology}

\begin{figure}
	\centering
	\includegraphics[scale=0.6]{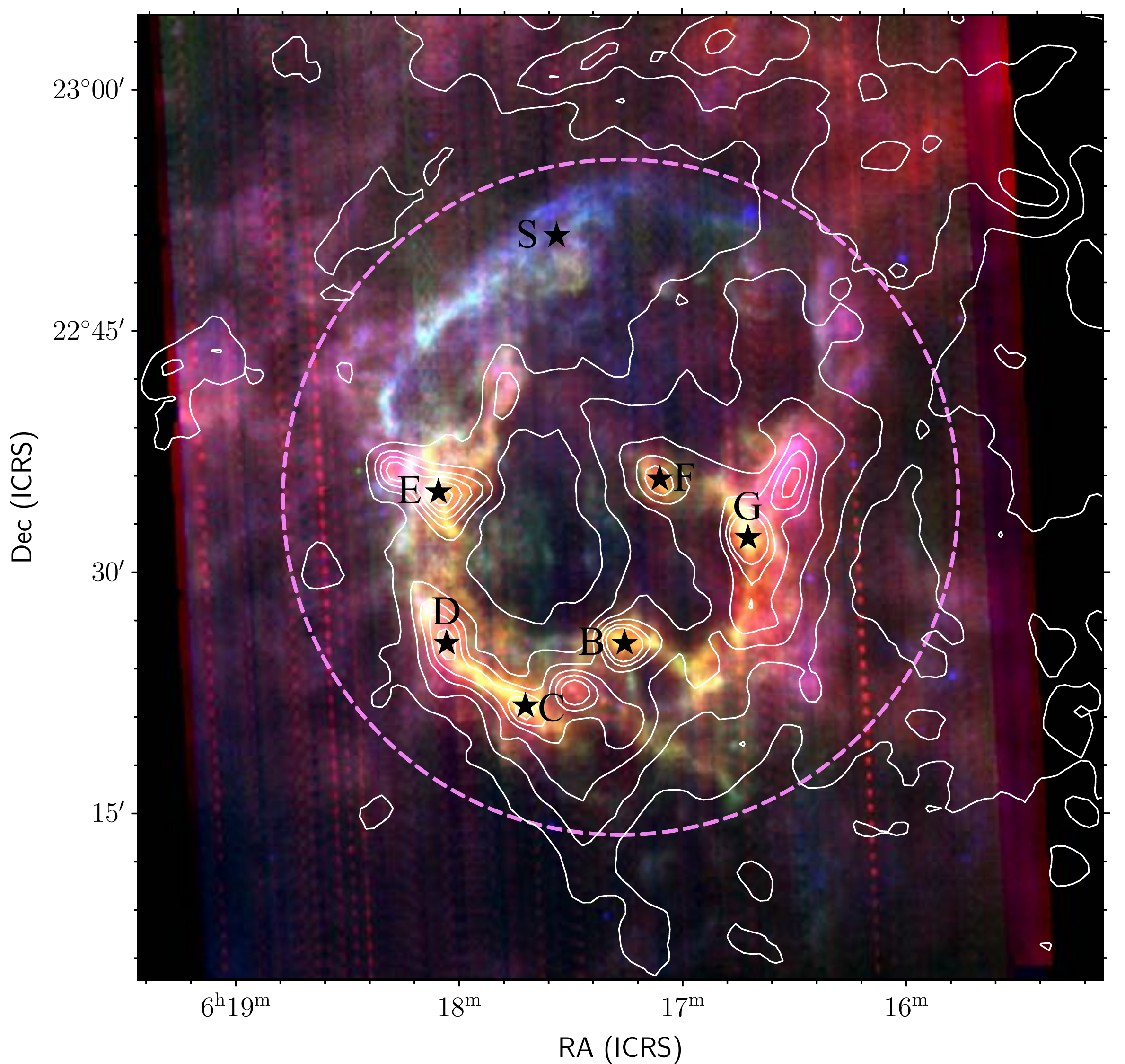}
	\caption{Three-color infrared image of \ic\ in the \spitzer/MIPS 24\mum\ (blue), 70\mum\ (green) and 160\mum\ (red) bands. The \CO(J$=$1--0) emission integrated from $-10<v<10$\kms\
	is overlaid by the white contours. The major molecular clumps B--G identified by \cite{Dickman1992} and a shell structure S relevant to the present study are labelled with black asterisks. The violet dashed circle at a radius of 21\arcmin\ indicates the photometry aperture.
	\label{fig:mips_rgb} }
\end{figure}

Figure \ref{fig:mips_rgb} displays the three-color image made from the \spitzer/MIPS infrared emission, i.e. at 24\mum\ in blue, 70\mum\ in green and 160\mum\ in red. The overall IR image is shell-like, specifically the 24\mum\ emission (blue) is brighter in the NE region, while the 160\mum\ (red) emission concentrates more in the southern region.
The NE region of \ic\ has multiple filaments, which is dominated by the 24\mum\ emission. It is suggested that the NE filaments represent shock-heated gas \citep{Fesen1984}. The atomic fine structure lines account for most of the 24\mum\ emission along the NE rim \citep{Noriega-Crespo2009,Rho2001}.
The southern ridge is composed of denser clumps and knots, which is part of the molecular cloud and where the infrared emission comes primarily from heated dust.
Various CO observations reveal that \ic\ is interacting with a molecular cloud that crosses the entire SNR from northwest to southeast \citep{Cornett1977,Su2014}. The $\CO (J=1-0)$ line emission at 115\ghz\ with a spatial resolution of $ \sim54''$ \citep{Su2014} observed by the 13.7\,m telescope of the Purple Mountain Observatory at Delingha is shown by white contours in Figure \ref{fig:mips_rgb} for comparison. The major molecular emission clumps B--G identified by \cite{Dickman1992} and a shell structure S relevant to the present study are labelled with black asterisks in this figure.

\subsubsection{Contributions from the Line and Synchrotron Emission}

The line emission as well as the dust radiation contributes to the IR fluxes. The distribution of [\mbox{O\,{\sc i}}] is similar to that of the shock-excited $\rm H_2$ and all the accelerated species, e.g. CO, $\rm HCO^+$ and HCN \citep{Burton1990}.
\cite{Oliva1999} suggested that most of the \iras\ 12\mum\ and 25\mum\ fluxes should be attributed to the ionized line emission (i.e. [\mbox{Ne\,{\sc ii}}] and [\mbox{Fe\,{\sc ii}}]), meanwhile they argued that the contamination of [\mbox{O\,{\sc ii}}] to the \iras\ 60\mum\ flux is unimportant in \ic. The \spitzer/IRS spectrum reported by \cite{Noriega-Crespo2009} also detected the [\mbox{Fe\,{\sc ii}}] line at 26\mum.
\cite{Mufson1986} estimated quantitatively the line contributions to the \iras\ 12, 25 and 60\mum\ fluxes are 90\%, 75\% and 37\%, respectively, by assuming a fixed $N_{\rm H}=10^{21}\,\rm cm^{-2}$. They argued that most of the \iras\ 100\mum\ fluxes are from radiatively heated dust rather than line emission. \cite{Burton1990} estimated that the [\mbox{O\,{\sc i}}] 63\mum\ line emission accounts for about $7\,\%$ of the \iras\ 60\mum\ emission for the entire SNR, but may contribute up to 40\% at the shock-peak positions. Taking the line emission into account, the measurements at $\lambda < 20\mum$, where the line emission dominates, are not considered during the model fitting. Besides, the fluxes at $\lambda > 20\mum$ are not corrected for the line emission because the warm dust is not the major component of the SNR and the amount of line contributions bears some uncertainty. Nevertheless, it will be discussed later the effect of subtracting line emission in the other bands.

In addition to the line emission at the shorter wavelength, the synchrotron emission at the longer wavelengths should be considered as well. For this, a simple power law of $S_\nu \propto \nu^{-\alpha}$ is adopted. It is found that the spectral index $\alpha$ varies across the SNR and with the frequency, specifically the NE region has a flatter spectral index, i.e. $0.05<\alpha<0.25$, while the diffuse interior region has a steeper $\alpha$ of $\sim 0.7$ \citep{Castelletti2011}. We take the average spectral index $\alpha$  being 0.36 and the flux density at 1\ghz\ (i.e. $S_{1\ghz}$) of 165\jy\ \citep{Green2019snr}. Then the synchrotron emission from mid-infrared to sub-millimeter wavebands are subtracted by extrapolating from the radio power-law emission. The synchrotron emission contribution decreases to only 5.3\% of the \planck\ 350\mum\ flux, and becomes negligible at the shorter wavelengths.

\subsubsection{The Global SED Fitting}\label{sec:sed_fit}

After subtracting the synchrotron emission and dropping the measurements at $\lambda < 20\mum$, the spectral energy distribution of the entire \ic\ is fitted by a modified two-component blackbody emission arising from the warm and cold dust respectively:
\begin{equation}\label{equ:sed_fit}
F_\nu(\lambda) = \frac{M_{\rm w}}{D^2} \kappa(a,\lambda) B_\nu(\lambda,T_{\rm w}) + \frac{M_{\rm c}}{D^2} \kappa(a,\lambda) B_\nu(\lambda,T_{\rm c})
\end{equation}
where (\mw, \tw) and (\mc, \tc) are the dust mass and temperature of the warm and cold component respectively, $B_\nu(\lambda,T)$ is the Planck function, $D$ is the distance to \ic\ and taken as 1.8\kpc. The mass absorption coefficient (also known as opacity) $\kappa(a,\lambda)$ (in units of $\rm cm^2\,g^{-1}$) is size-dependent and calculated from the dust emissivity $Q(a,\lambda)$ and grain mass density $\rho$ by $\kappa(a,\lambda)=3Q(a,\lambda)/4\rho a$. We adopt a silicate and graphite mixture model with the dielectric constants from \cite{Draine1984} and the mass ratio fixed at $M_{\rm sil}/M_{\rm gra}=2:1$. Actually $\kappa$ for both silicate and graphite is nearly independent of grain radius ($a$) at wavelengths longer than $10\mum$ for grain size $<1\mum$, therefore a typical grain size of $a=0.1\mum$ is assumed in Equation \ref{equ:sed_fit}. The data at the \planck\ 1380\mum\ is excluded in the SED fitting because the  wavelength range of dust optical constant is unavailable  at this wavelength.

\begin{figure}
	\centering
	\includegraphics[scale=0.8]{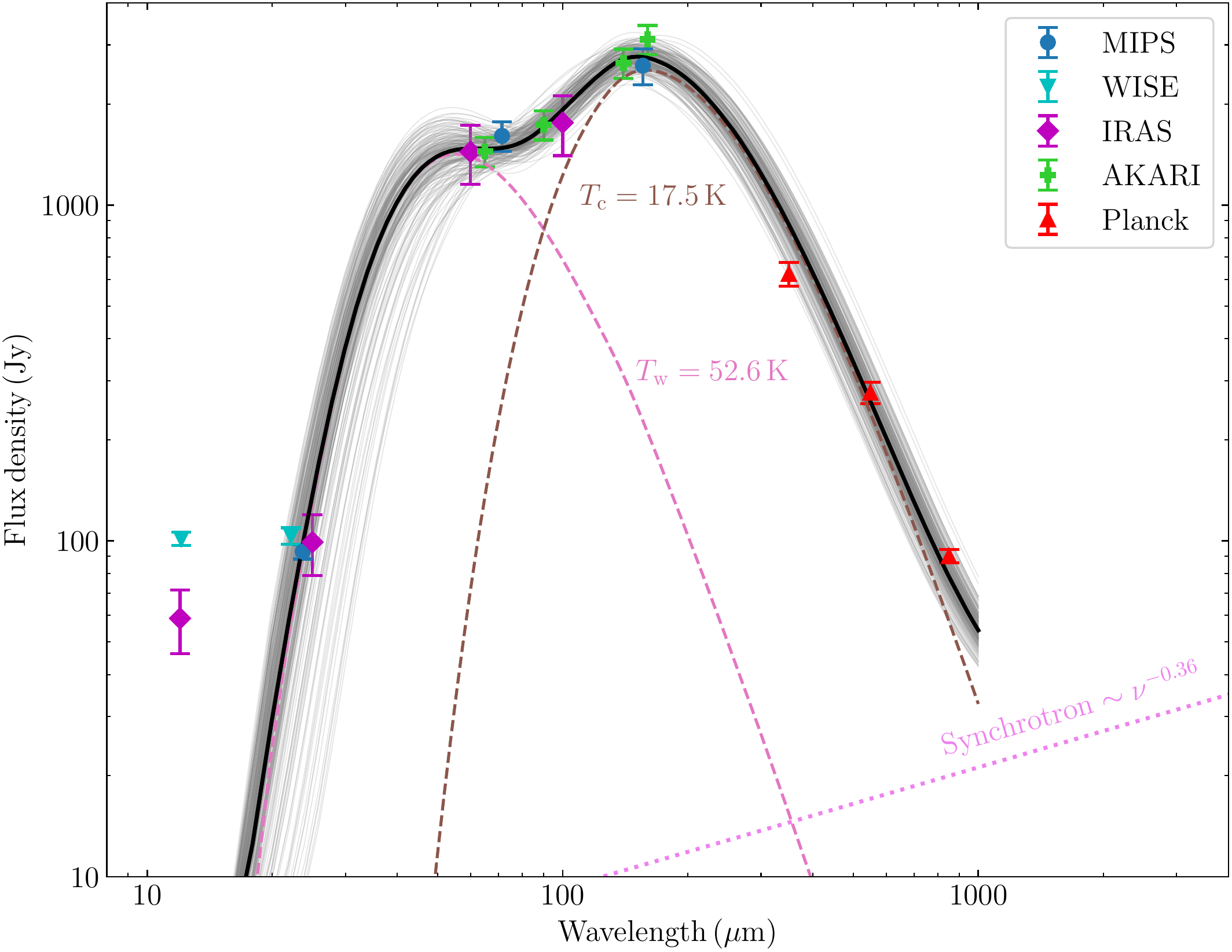}
	\caption{The global spectral energy distribution of \ic\ from IR to submm assembled from the measurements by \spitzer/MIPS, \wise, \iras, \akari\ and \planck. The points with $1\sigma$ error bar represent the photometry data without subtracting the contribution from line emission. The fitting is made to the SED at $\lambda>20$\mum\ by using a two-component model, where the best-fitted warm ($\tw=52.6$\K) and cold ($\tc=17.5$\K) components are shown by the pink and brown dashed line respectively, and the black solid line represents the sum of the two dust components and synchrotron emission (the violet dotted line). The translucent grey lines show 200 SEDs that are calculated by the parameters randomly selected from the bootstrapped distributions.
	\label{fig:sed_posterior} }
\end{figure}

\begin{figure}
	\centering
	\includegraphics[scale=0.6]{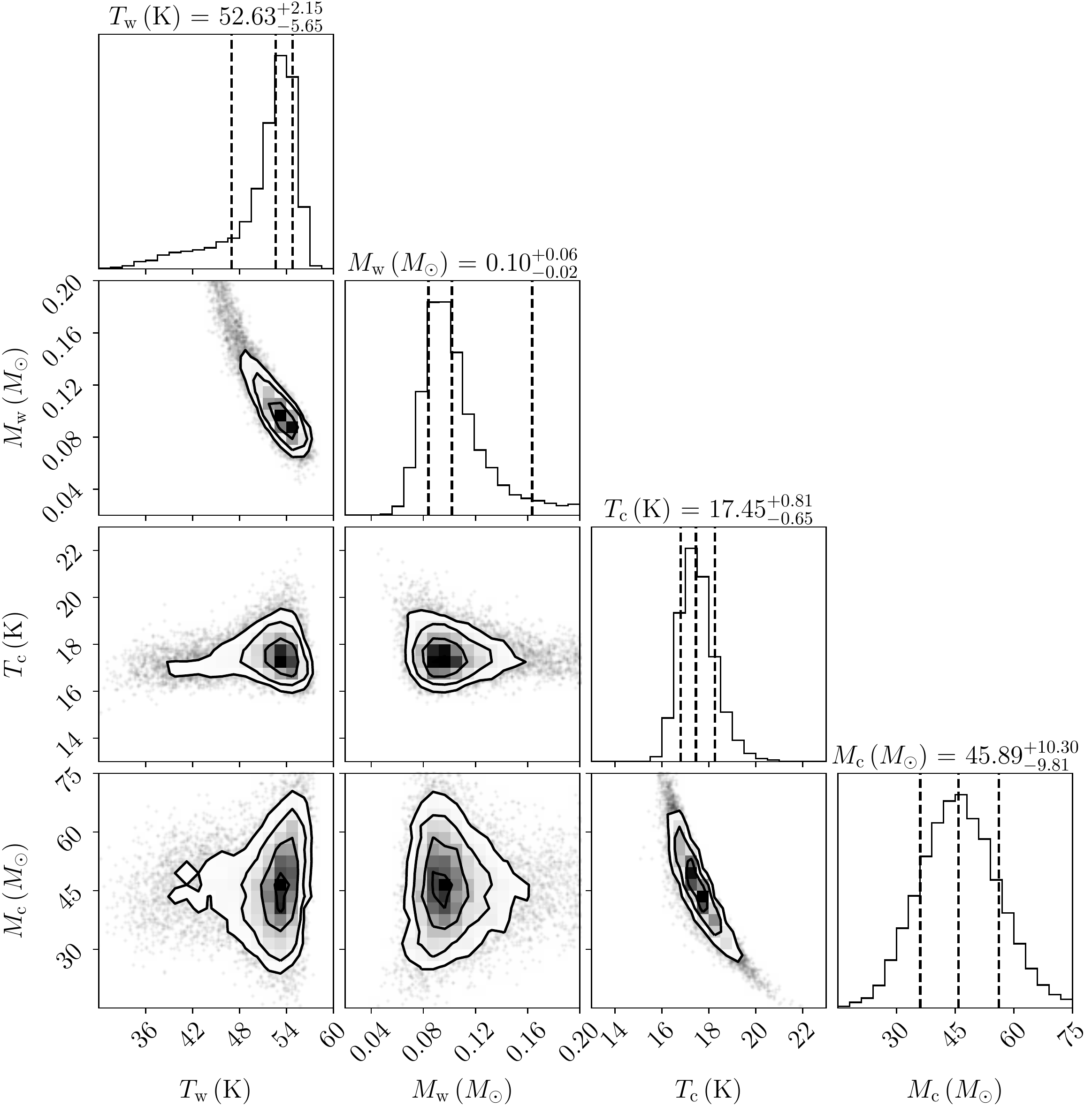}
	\caption{The probability distribution of the parameters for the model fitting to the global infrared SED shown in Figure \ref{fig:sed_posterior}. The dashed lines show the best-fit solution and $1\sigma$ interval for the whole range.
	\label{fig:sed_corner} }
\end{figure}

In order to calculate the uncertainties of the parameters (\mw, \tw, \mc, \tc), a bootstrap technique is used by fitting 2000 randomly sampled SEDs based on a Gaussian distribution with the photometry uncertainty as the dispersion.
The result of fitting the global SED is displayed in Figure \ref{fig:sed_posterior}, and
Figure \ref{fig:sed_corner} presents the probability distribution of the best-fit values of the parameters. The parameters' uncertainties are assigned by the 68\% (i.e. $1\sigma$) interval in both the upper and lower parameter distributions, which are generally asymmetric.
The bootstrap technique yields a warm dust component with the temperature of $52.6^{+2.2}_{-5.7}$\K\ and the mass of $0.10^{+0.06}_{-0.02}$\msun, and another cold dust component with the temperature of $17.5^{+0.8}_{-0.7}$\K\ and the mass of $45.9^{+10.3}_{-9.8}$\msun, respectively. The temperature of the warm dust, \tw, agrees very well with the result of \citet{Kokusho2013}, while the temperature of the cold component is higher than that of \citet{Kokusho2013}. Meanwhile, \tw\ is higher than the temperature of the shock-headed dust from fitting the IRAS photometry by \citet{Mufson1986}.

If the line contribution is subtracted from the fluxes by 75\% in the \wise\ 22\mum, MIPS 24\mum\ and \iras\ 25\mum\ bands, 40\% in the MIPS 70\mum, \iras\ 60\mum\ and \akari\ 65\mum\ bands, the warm dust temperature and mass would become $49.7^{+4.3}_{-14.9}$\K\ and $0.07^{+0.37}_{-0.03}$\msun, respectively. The warm temperature is decreased by 5.3\% and the warm mass is decreased by 30.0\%. The temperature of cold dust is little changed, becoming $18.2^{+1.1}_{-0.9}$\K\ and the mass of cold component becomes $37.7^{+9.3}_{-8.9}$\msun\ that is slightly smaller, which is expected since the cold dust emits mainly at longer wavelength. This difference may be kept in mind in further analysis which does not consider the line emission at $\lambda > 20$\mum.

The mass of clod dust is more than two orders of magnitude larger than that of the warm dust. The warm component is certainly heated by the SN shocks and traces the dust associated with \ic, while the cold component may be dominated by the dust in the foreground and/or background molecular cloud. Because \ic\ is located within the molecular cloud, it is difficult to completely decompose the SNR from the surrounding cloud. The later analysis of the map of dust temperature would reveal more details.

\subsubsection{The Spatially Resolved Fitting: Maps of Dust Temperature and Mass}\label{sec:dust_map}

\begin{figure*}
	\centering
	\includegraphics[scale=0.5]{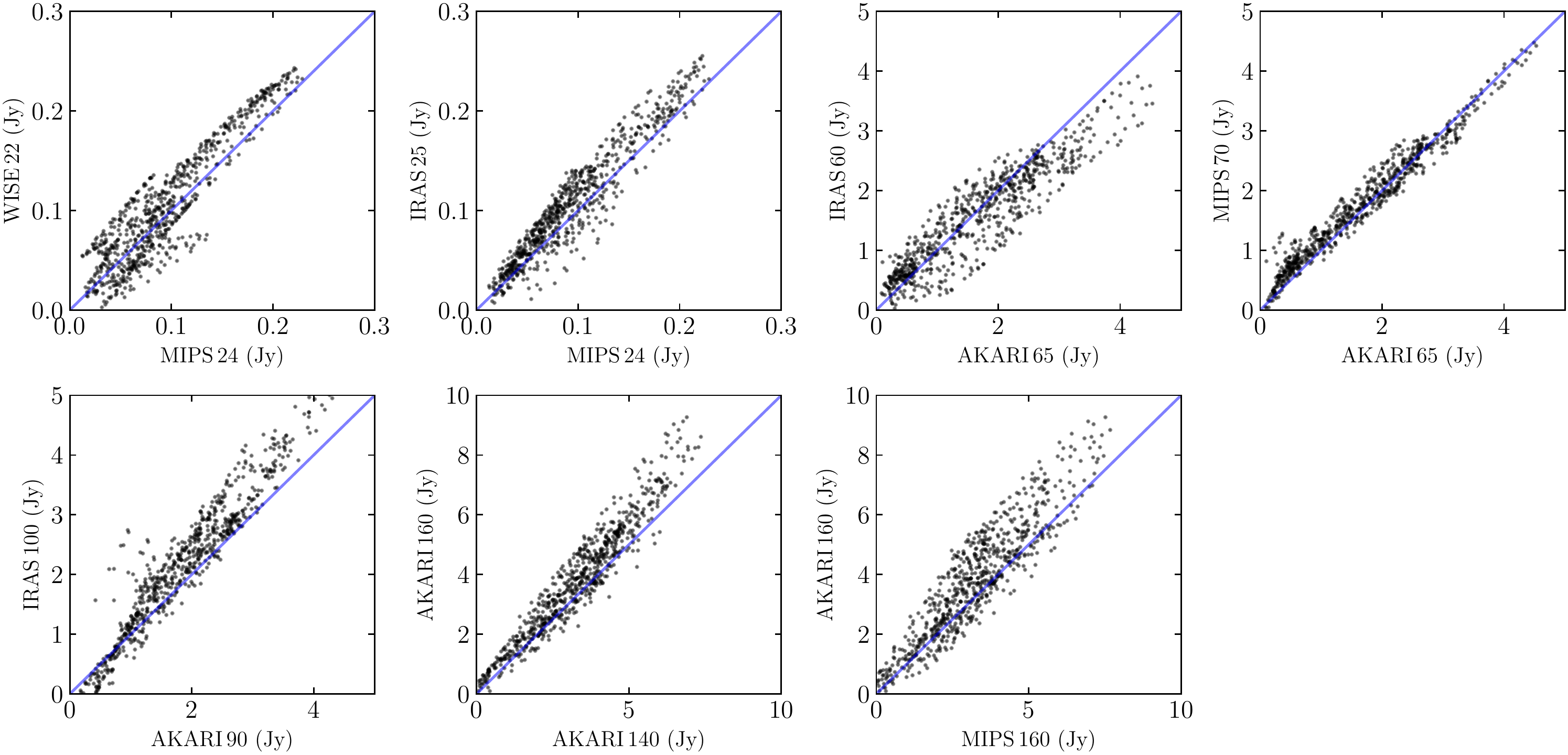}
	\caption{The comparison for the pixel intensity (in units of Jy/pixel) in two similar bands. All maps are convolved to the same spatial resolution of 300\arcsec\ and regridded to the same pixel scale of 60\arcsec. The blue solid lines show the one-to-one correspondence.
	\label{fig:pix_compare} }
\end{figure*}

The spatial variation of dust properties (\mw, \tw, \mc, \tc) in \ic\ can be investigated by fitting the SED pixel-by-pixel. All the images after subtracting the background emission are convolved to the \iras\ 100\mum\ resolution of 300\arcsec\ using a Gaussian kernel, then regridded to a pixel size of 60\arcsec. In order to obtain reliable results, the pixel-by-pixel fitting requires all fluxes of individual pixel to be positive. After the bad pixels are masked, a total of 720 pixels are reserved. The pixel intensity in two similar bands are compared in Figure \ref{fig:pix_compare}, which agrees highly with each other and confirms the convolution correct.

\begin{figure}
	\centering
	\includegraphics[scale=0.6]{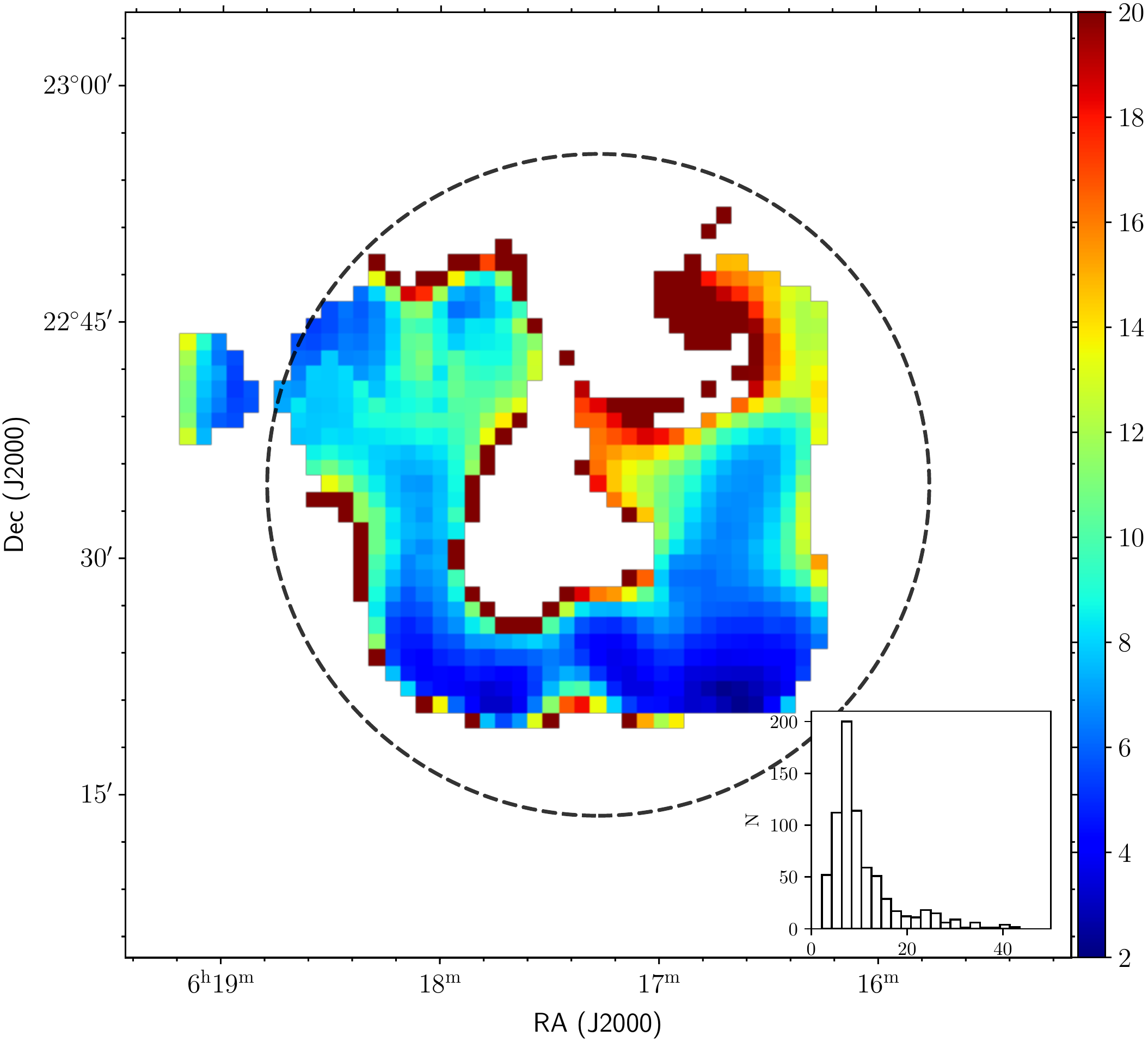}
	\caption{The reduced $\chi^2$ distribution derived from fitting the MIR to submm SEDs of \ic\ pixel-by-pixel. The inset illustrates the histogram of reduced $\chi^2$. The median reduced $\chi^2\approx 8.4$. The black dashed circle indicates the photometry aperture radius at 21\arcmin.
	\label{fig:chi2_map}}
\end{figure}

\begin{figure*}
	\centering
	\includegraphics[scale=0.5]{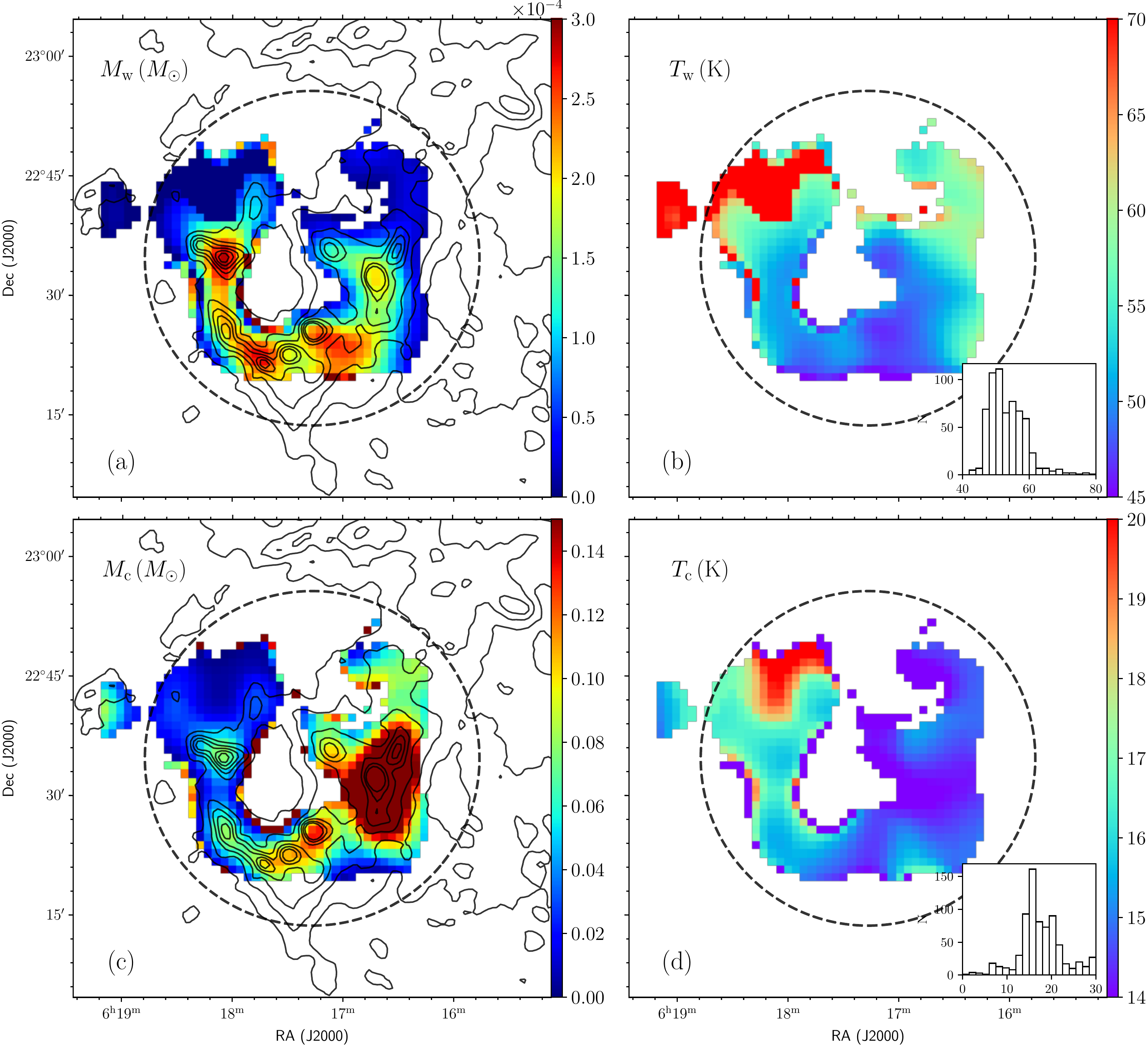}
	\caption{The dust mass map in unit of \msun/pixel and the temperature map in unit of Kelvin derived from a two-component model fitting the mid-IR to submm SED of \ic: the mass (a) and temperature (b) of warm dust, and the mass (c) and temperature (d) of cold dust. Panels (a) and (c) are overlaid with the integrated \CO\ emission contours for comparison. The black dashed circle (radius of 21\arcmin) indicates the photometry circle. The insets in panels (b) and (d) show the histogram of temperature with the median value of 53.9\K\ and 16.7\K\ respectively.
	\label{fig:temp_mass_maps}}
\end{figure*}

Since the error of the integrated fluxes at $\lambda \leq 160\mum$ is dominated by  the calibration uncertainty, the flux error of individual pixel is assigned to the corresponding calibration uncertainty. For the \planck\ bands, the relative errors take around 8.2\%, 7.6\% and 4.7\%  at 350\mum, 550\mum\ and 850\mum, respectively. 
Using the same model as in the global fitting, the reduced $\chi^2$ for each pixel is calculated by:
\begin{equation}
\frac{\chi^2}{\rm dof}=\frac{\sum_{j=1}^{N_{\rm obs}}\left[F_\nu(\lambda_j)^{\rm mod} - F_\nu(\lambda_j)^{\rm obs} \right ]^2/\sigma(\lambda_j)^2}{N_{\rm obs}-N_{\rm para}}
    \end{equation}
where $F_\nu(\lambda_j)^{\rm mod}$ and $F_\nu(\lambda_j)^{\rm obs}$ are the modelled and observed pixel fluxes respectively, $\sigma(\lambda_j)$ is the uncertainty of individual pixel flux as calculated above, $N_{\rm obs}=14$ and $N_{\rm para}=4$ are the number of observational data and model parameter respectively. The $\chi^2/{\rm dof}$ map of the pixel-by-pixel fitting is shown in Figure \ref{fig:chi2_map}.
The median of $\chi^2/{\rm dof}$ is $\sim8.4$, and some pixels with $\chi^2/{\rm dof} >20$ are located at the inner and outer edge of the SNR shell. The resultant distribution of dust mass and temperature of the warm and cold component is displayed in Figure \ref{fig:temp_mass_maps}. It can be seen that both the warm and cold dust mainly concentrate along the shell, and there is a massive clump in the west of the remnant, which accounts for a half of the dust mass (i.e. $\sim$24\msun) to the SNR. This clump position agrees with clump G of the CO line emission identified by \citet{Dickman1992} (c.f. Figure \ref{fig:mips_rgb}). From the comparison between the \CO\ emission and the mass distributions in Figure \ref{fig:temp_mass_maps} (a) and (c), it can be seen that the cold clumps coincide with the molecular clumps better than the warm clumps. The spatial distributions of the warm and cold components show distinction in the southern rim of the shell where most of the shocked H\,{\sc i} gas and molecular emission is located \citep{Lee2008,Lee2012,Burton1988}, suggesting the ongoing interaction between the SNR and the molecular clouds.

The temperature of the warm component varies mainly from 40--60\K, and the cold component from 12--25\K. In addition, the temperature of both the warm and cold components is relatively higher in the NE region than in other regions. Some pixels of the inner edge have very low temperature (i.e. $<$10\K) in Figure \ref{fig:temp_mass_maps} (d), which may not be true but caused by the large uncertainty due to the low S/N (cf. Figure \ref{fig:chi2_map}). As shown in Figure \ref{fig:temp_mass_maps} (b), the temperature distribution of warm component appears relatively uniform around 50\K\ along the shell, suggesting that the warm component is likely associated with the shock-heated dust by \ic. While for the cold component, the temperature in the west is lower than in the east (cf. Figure \ref{fig:temp_mass_maps} (d)). A large portion of the dust mass in the cold component may come from the ambient molecular cloud. When a one-temperature model is taken to fit the SED of the obvious MC in the SNR's northwest, the dust temperature is found to be a few K lower than the SNR. This indicates that the dust within the SNR area is at least partially heated by the SN explosion. But it's true that some of the MC region may be intact, in particular in the western part.

\section{Dust Mass Derived from the Extinction} \label{sec:extinction}

\subsection{Identification of the SNR Extinction Component}

The extinction of \ic\ has been studied by a few works, e.g. \citet{Yu2019} and \citet{Zhao2020}. In order to identify the extinction caused by \ic,  we choose the 3D reddening map based on \emph{Gaia}, Pan-STARRS1 and 2MASS by \cite{Green2019} for its more complete coverage in both distance and area.  The dust map is retrieved  by the BayestarQuery function in the Python package dustmaps\footnote{https://dustmaps.readthedocs.io/}, which has a scale of 3.4\arcmin\ on a voxel side.

As a first step, the variation of extinction with the distance is retrieved for several representative positions identified in the molecular emission map labeled as clumps B-G  by \citet{Dickman1992} in Figure \ref{fig:mips_rgb}. We also select a shell clump (named ``S'') located in the NE rim as marked in Figure \ref{fig:mips_rgb}, which shows strong IR emission but no CO emission.

\begin{figure}
	\centering
	\includegraphics[scale=0.7]{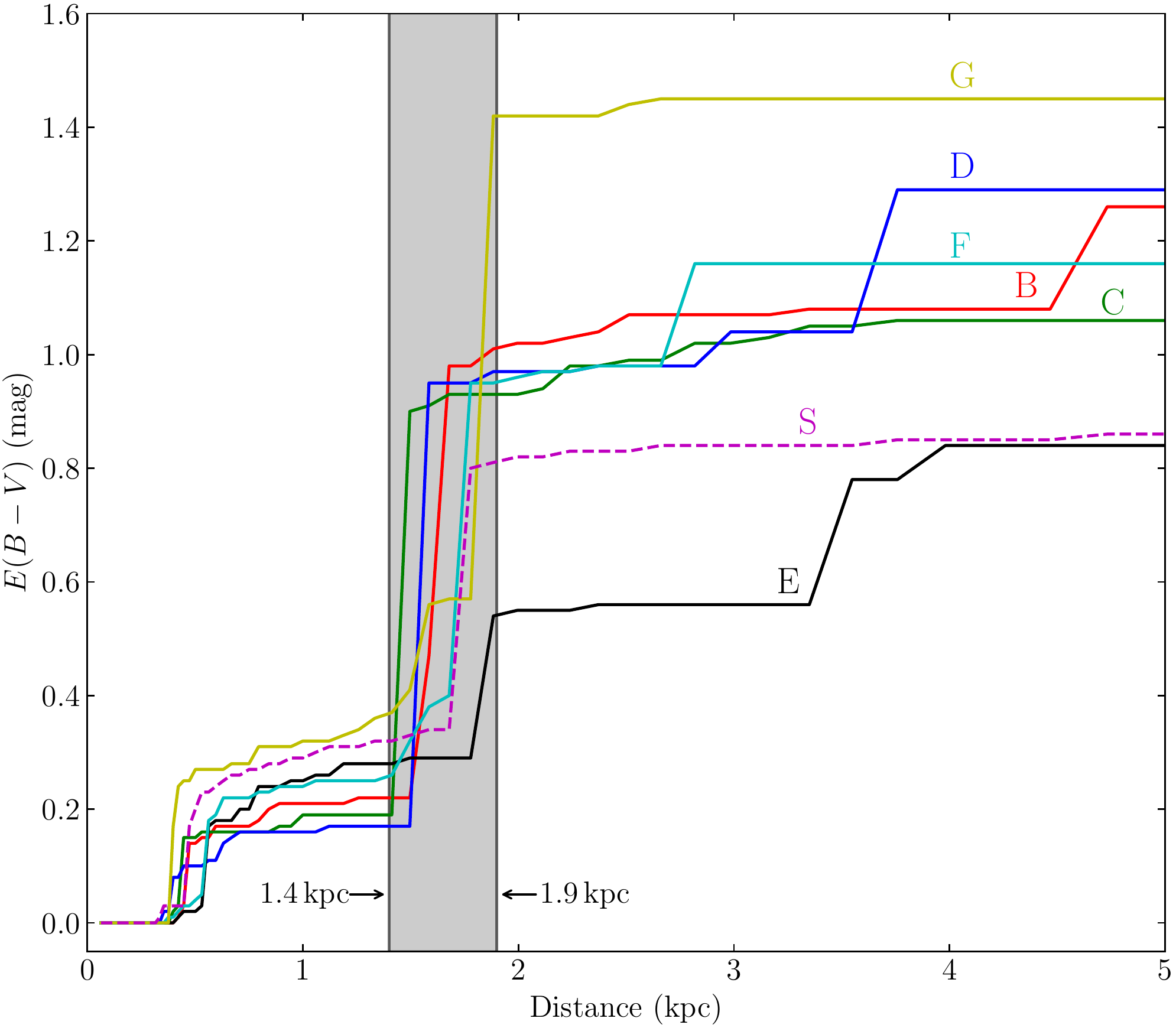}
	\caption{The color excess \EBV\ (magnitude) as a function of distance (kpc) extracted from the 3D extinction map of \cite{Green2019}. The lines in color correspond to the 7 locations marked in Figure \ref{fig:mips_rgb}. The vertical grey bar indicates the extinction jumps toward \ic\ at the distance range of 1.4--1.9\kpc.
	\label{fig:ext_to_dist} }
\end{figure}

Figure \ref{fig:ext_to_dist} displays the variation of the color excess \EBV\ with the distance for the representative positions.  These positions exhibit a few common jumps: one of $\sim 0.2$\,mag at 0.3--0.5\kpc, one of $\sim 0.7$\,mag at 1.4--1.9\kpc, and a few other small ones at different distance behind 2.0\kpc.  According to previous studies, the distance of \ic\ is around 1.8\kpc, then the second jump is consistent with  the SNR. As the above analysis of infrared radiation shows that some cold dust may come from the intact ambient molecular cloud, the extinction can also partly originate from them.

\begin{figure*}
    \centering
	\includegraphics[scale=0.7]{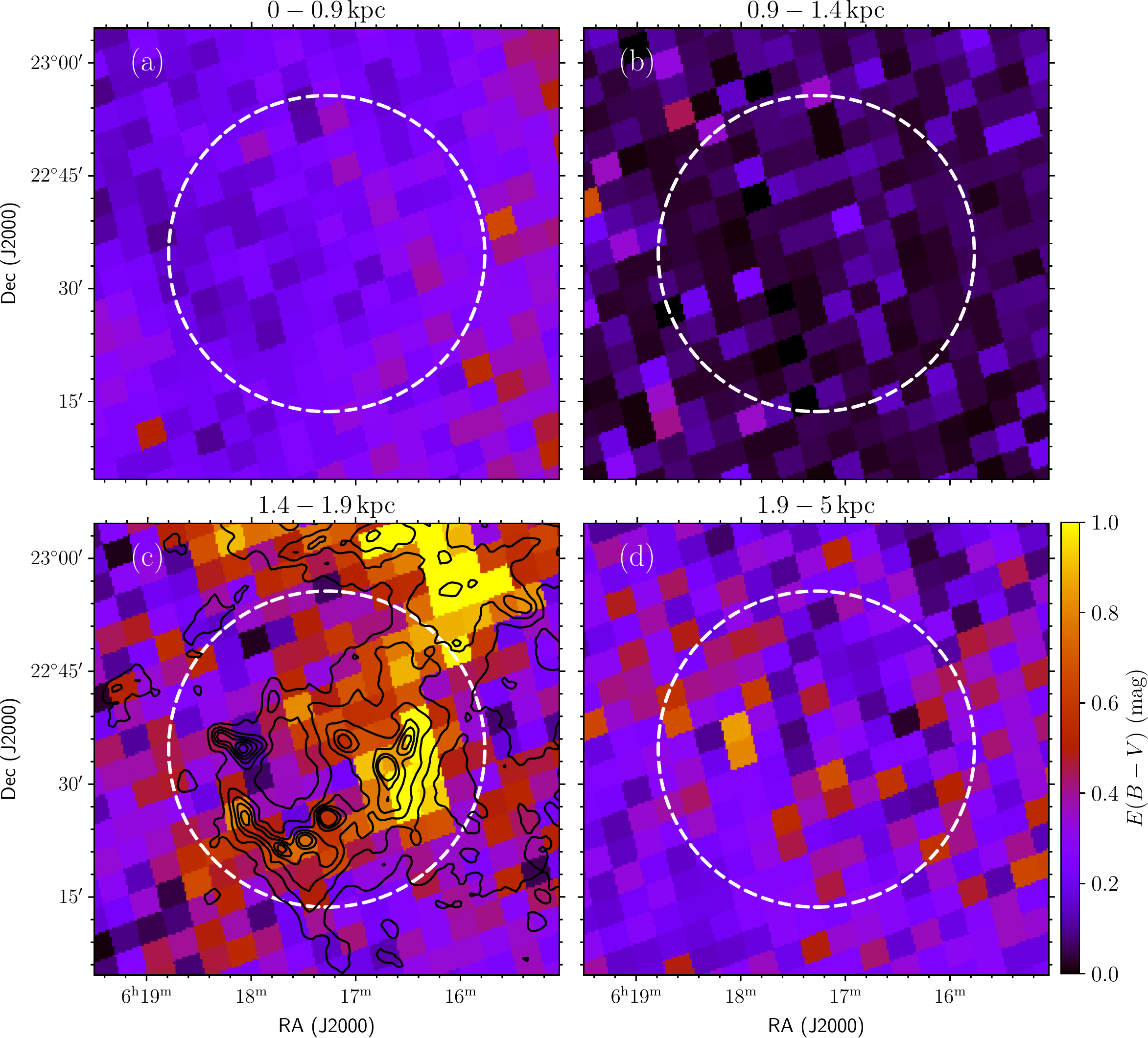}
	\caption{The Bayestar 3D extinction map \citep{Green2019} in the sightline of \ic\ at the distance bin of 0--0.9\kpc\ (a), 0.9--1.4\kpc\ (b), 1.4--1.9\kpc\ (c) and 1.9--5.0\kpc\ (d), overlaid with the integrated \CO\ emission contours (c). The position of photometry aperture is shown with a white dashed circle. The color bar marks the value of \EBV\ in the range of 0--1.0\,mag.
	\label{fig:3d_maps} }
\end{figure*}

We divide the distance into four parts: 0--0.9\kpc, 0.9--1.4\kpc, 1.4--1.9\kpc, and 1.9--5\kpc. The extinction map of each part is displayed in Figure \ref{fig:3d_maps}. It is apparent that the extinction map from $d=$1.4--1.9\kpc\ exhibits the structure of the SNR and coincides with the dust and \CO\ emission. The cloud at 0.3--0.5\kpc\ is a foreground cloud with no similarity to the structure of \ic\ at all. The maps for the distance in the range of 0--0.9\kpc\ and 0.9--1.4\kpc\ show the feature of diffuse medium with no visible structure, and for the distance behind 1.9\kpc\, the extinction results from possibly a few background clouds. It should be mentioned that the position ``S" behaves similar to others, which indicates that the dust there is part of the SNR. Though the dispersion in the distance of various clumps may come from the structure in depth, it is more possibly caused by the uncertainty of distance. In conclusion, the extinction in the range of 1.4--1.9\kpc\ is produced by dust associated with \ic\ and the surrounding molecular cloud.

The extinction map of \ic\ is constructed after the extinction component is identified. Specifically, the increase of extinction  from $d=1.4$\kpc\ to $d=1.9$\kpc\ is taken as the extinction of \ic\ for each pixel, which forms the extinction map in Figure \ref{fig:3d_maps} (c).  The amount of \EBV\ ranges around 0.3--1.5\,mag, which converts to the visual extinction of \AV$\sim$ 1--5\,mag. The extinction  specifically, the color excess \EBV\ at clump G has the largest extinction, which is about 1 mag, inferred \AV$\sim$3--4\,mag depending the \RV\ value. This agrees with the result of \cite{Dishoeck1993} who reported the optical extinction to be \AV$\sim$3--4$\,\rm mag$ derived from the CO observation. On the other hand, \citet{Richter1995} reported a much larger extinction of $\AV=10.8\,\rm mag$ at clump G, and also found a large optical extinctions of $\AV=13.5\,\rm mag$ for clumps B and C by measuring the $\rm H_2$ line ratios. Such discrepancy may be induced from the gas-to-dust ratio, or different spatial resolution for the line and extinction observations. The dust mass calculated from the infrared emission in Section \ref{sec:dust_map} shows that the mass of position G is also the largest, which agrees with the largest extinction.

\subsection{The Dust Mass from Extinction}

The dust mass of a SNR can be estimated from its extinction because the extinction is proportional to the dust column density. The details of such method are described in \citet{Zhao2018}. In short, the dust mass responsible for the extinction is:
\begin{equation}\label{equ:mass_ext}
M_{\rm d}^{\rm {ext}}=\Sigma_{\rm dust} \times A_{\rm SNR}=\frac{\AV}{\Kextv}\times A_{\rm SNR}
\end{equation}
where $\Sigma_{\rm dust}$ is the  mass surface density, \Kextv\ is the mass extinction coefficient for the $V$ band in the unit of $\rm{mag\,cm^2\,g^{-1}}$, and $A_{\rm SNR}$ is the area of the SNR.

The parameter \Kextv\ depends on the dust model that relates mainly to  dust components and size distribution. We calculate this parameter with the dust model which is constructed by fitting the average extinction curve of \ic\ based on the spectroscopic and multi-band photometric data  by \citet{Zhao2020}. This dust model consists of mixed silicate and graphite grains with a mass ratio of $M_{\rm sil}/M_{\rm gra}= 2:1$, which is the same as used in estimating the dust mass from infrared emission in
Section \ref{sec:sed_fit}. The dust grain size distribution follows the MRN power-law with a slope of $\alpha$ and a range in size from 0.005 to 0.25\mum\ \citep{Mathis1977}. Again, the bootstrap method is used to estimate the uncertainty of the mass extinction coefficient from fitting the extinction curve of \ic. This modeling results in the mass extinction coefficient $\Kextv=(3.7^{+0.5}_{-0.8})\times 10^4\,{\rm mag\,cm^{2}\,g^{-1}}$, corresponding to $7.8^{+1.1}_{-1.6}\,{\rm mag\,pc^{2}}\msun^{-1}$. The value of \Kextv\ is almost the same as $\Kextv=(3.7\pm0.5)\times 10^4\,{\rm mag\,cm^{2}\,g^{-1}}$ derived by \citet{Nozawa2013},  while about 30\% larger than $2.8\times 10^4\,{\rm mag\,cm^{2}\,g^{-1}}$ based on the model of \citet{Weingartner2001}. The value of $\AV$ is obtained from converting the Bayestar reddening \EBV\ from \cite{Green2019} with the recommended coefficient $\AV=2.742\times\EBV$ \citep{Schlafly2011}. The area of \ic\ is calculated with the distance of 1.8\kpc\ and an angular diameter of 21$'$.

By summing up the extinction in the SNR distance range of 1.4--1.9\kpc\ over the area within the 21\arcmin\ circle, the dust mass is calculated to be about 
$M_{\rm d}^{\rm {ext}}=66.0_{-8.1}^{+17.1}\msun$
with $K_V^{\rm {ext}}=3.7\times 10^4\,{\rm mag\,cm^{2}\,g^{-1}}$. According to  the average extinction of \ic\ to be $\AV=1.0\,\rm mag$ estimated by \citet{Zhao2020} based on fewer stars with spectroscopically determined stellar parameters,  the dust mass would be $M_{\rm d}^{\rm {ext}}=\AV / K_V^{\rm {ext}}\times \pi R_{\rm SNR}^2\approx 50.4_{-6.2}^{+13.0}\msun$ if assuming a uniform extinction distribution within the SNR. Indeed, this simple approach results in a roughly consistent estimation with that derived from  the spatial integration.

\section{Unification of the Emission Mass and the Extinction Mass} \label{sec:unification}

The above calculations yield a mass of $M_{\rm d}^{\rm emi}=45.9^{+10.3}_{-9.8}$\msun\ from the emission and $M_{\rm d}^{\rm {ext}}=66.0_{-8.1}^{+17.1}$\msun\ from the spatially-integrated extinction of the SNR \ic. These two results agree with each other within the large uncertainties, though the mass from extinction, $M_{\rm d}^{\rm ext}$ is about 40\% larger than that from emission $M_{\rm d}^{\rm emi}$. The distance may suffer some uncertainty, but it affects $M_{\rm d}^{\rm ext}$ and $M_{\rm d}^{\rm emi}$ in the same way  so that its effect is cancelled out. It deserves to mention that the mass is proportional to the distance squared. If a distance of 1.5\kpc\ is adopted, both $M_{\rm d}^{\rm ext}$ and $M_{\rm d}^{\rm emi}$ would decrease by 30\%. On the side of $M_{\rm d}^{\rm {ext}}$, it could be overestimated due to the integration of extinction over a wide distance range of 500\pc, which is larger than the typical size of a giant molecular cloud (i.e. $<200$\pc).  On the other hand, the mass extinction coefficient \Kextv\ we adopted is moderate.  If the \Kextv\ value of \cite{Weingartner2001} is used, the extinction mass would be raised to $M_{\rm d}^{\rm ext}\approx90\msun$, much more higher than $M_{\rm d}^{\rm emi}$.

On the side of $M_{\rm d}^{\rm emi}$, it may also take different values.
The dust mass estimated from fitting the SED is highly dependent on the mass absorption coefficients $\kappa(\lambda)$. In general, $\kappa(\lambda)$ is approximated by a single power law:
\begin{equation}
\kappa(\lambda) = \kappa_{\lambda_0} \left( \frac{\lambda_0}{\lambda}\right )^\beta
\end{equation}
where $\kappa_{\lambda_0}$ is the $\kappa$ value at reference wavelength $\lambda_0$ and $\beta$ is the dust emissivity index. In the dust model for the above calculation in Section \ref{sec:sed_fit}, the opacity of \cite{Draine1984} is adopted, which calculates the opacity at $\lambda>100$\mum\ by extrapolating the experimental data at $\lambda <100$\mum, yielding $\kappa$ at $500\mum$ (i.e. $\kappa_{500}$) of 1.17 and 2.0\,$\rm cm^2\,g^{-1}$, and $\beta$ at $\lambda > 50 \mum$ of 2.1 and 2.0 for silicate and graphite respectively. For the mixed dust model with $M_{\rm sil}/M_{\rm gra}=2:1$, the compromised $\kappa_{500}=1.45\rm\,cm^2\,g^{-1}$ and $\beta=2.06$. However, some experimental and observational studies find that $\kappa(\lambda)$ may take other values by an order of magnitude difference depending on dust temperature and deviate from a power law \citep{Clark2016,Fanciullo2020}. Table \ref{tab:kappa} lists a few popular dust mass absorption coefficients $\kappa_{\lambda_0}$, which are used to fit the global SED of \ic. By leaving $\beta$ as a free parameter, we first derive the best-fit $\beta=1.53$, $\tw=55.1\K$ and $\tc=20.8\K$. For the given $\kappa_{\lambda_0}$ value of \citet{Hildebrand1983}, \citet{James2002} and \citet{Clark2016}, the estimated cold dust mass is 19.6\msun, 28.5\msun\ and 88.6\msun, respectively. For comparison, the dust mass is also calculated with the dust mass absorption coefficients of \citet{Li2001} which has approximately $\kappa_{500}=0.95\rm\,cm^2\,g^{-1}$ and $\beta=2.13$ and yields the cold dust mass of 74.2\msun. Apparently, the range of $M_{\rm d}^{\rm emi}$ obtained by various dust mass absorption coefficient, i.e. $\sim$ 20--90\msun, covers $M_{\rm d}^{\rm ext}$. Because the extinction mass is calculated from the extinction and the mass extinction coefficient in the visual band, it hardly depends on the far-infrared dust emissivity, and thus these alternative absorption coefficients for fitting the infrared SED does not change the extinction mass. This means that the extinction mass and the emission mass can be tuned to be the same if appropriate absorption coefficients are taken. In another words, the absorption coefficients of the dust in \ic\ can be better described by an intermediate value of \citet{Draine1984} and \citet{Clark2016} from the perspective of unifying the emission mass with the extinction mass.

In addition, optically thin radiation is assumed to calculate  $M_{\rm d}^{\rm emi}$ in Section \ref{sec:sed_fit}. This assumption should be true for the warm dust, but may not be true for the cold dust. The extinction map shows that the largest extinction at clump G reaches about \AV$\sim$3--4\,mag, apparently optically thick. The optically thin assumption then under-estimates the dust mass, consistent with the result that $M_{\rm d}^{\rm emi} < M_{\rm d}^{\rm ext}$. With the shortcoming of the optically thin assumption, we may conclude that $M_{\rm d}^{\rm ext}$ derived from the extinction can be more true, i.e. the dust mass of \ic\ is about 66.0\msun.


\section{Summary}
\label{sec:conclusion}

This work makes the first attempt to estimate the dust mass of a supernova remnant by bringing about a unified result from dust emission and extinction. The object \ic\ is chosen for its abundant observations of the dust emission in the infrared bands and the extinction in the visual bands. The dust mass is firstly estimated to be about 46\msun\ from fitting the spectral energy distribution in the mid-to-far infrared  bands observed by \spitzer/MIPS, \wise, \iras, \akari\ and \planck, which spans the range from about 10\mum\ to 1000\mum. The extinction caused by \ic\ is then determined by identifying the component which agrees with the spatial structure of the warm dust emission, i.e. at the distance around 1.8\kpc. According to the extinction map provided by the Bayestar, the dust mass is calculated to be about 66\msun\ with the dust mass absorption coefficient derived from the extinction curve. The dust masses based on the infrared emission and the visual extinction coincide within the uncertainty. On the other hand, the dust mass derived from infrared emission can differ by a factor of two to three with other dust opacities, and could be under-estimated due to the optically thin assumption for the calculation. The unification of emission and extinction mass sets some constraint on the dust properties. In conclusion, the dust mass derived from the extinction may be more reliable for \ic.


\begin{acknowledgments}
We would like to thank the anonymous referee for his/her insightful comments. We thank Professors Jian Gao and Haibo Yuan for very helpful suggestions and comments. We also thank Dr Yang Su for kindly providing the CO data. J.L. and H.Z. wish to acknowledge Drs. Mikako Matsuura and Phil Cigan for the help in teaching infrared photometry at Cardiff University. This work is supported by NSFC 12133002 and 11533002, National Key R\&D Program of China No. 2019YFA0405503 and CMS-CSST-2021-A09.

\end{acknowledgments}

%

\vspace{5mm}
\facilities{\emph{Spitzer}/MIPS, \wise, \iras, \akari, \planck}


\software{Astropy \citep{astropy2013},
        APLpy \citep{aplpy2012},
        dustmaps \citep{Green2018},
        Photutils \citep{photutils2020}.
          }






\bibliography{IC443}{}
\bibliographystyle{aasjournal}

\begin{table*}
\centering
\caption{The Multi-band Infrared Photometry of \ic}
\label{tab:data}
\begin{tabular}{c c c c c c c}
\hline \hline
Instrument & Band & Pixel Size & Resolution & Calibration &\multicolumn{2}{c}{Flux Density (Jy)} \\
  &  ($\rm \mu m$) &  (\arcsec) &  (\arcsec) &  Uncertainty & This Work & Published \\
\hline
\wise\ & 3.4 & 1.375 & 6.1 & 2.9\%  & $19.7\,\pm\,0.6$ & - \\
\wise\ & 4.6 & 1.375 & 6.4 & 3.4\%  & $23.8\,\pm\,0.8$ & - \\
\wise\ & 12  & 1.375 & 6.5 & 4.6\%  & $101.5\,\pm\,4.7$ & - \\
\wise\ & 22  & 1.375 & 12  & 5.6\%  & $103.6\,\pm\,5.8$ &- \\
\spitzer/MIPS & 24  & 2.45  & 6   & 5\%  & $92.9\,\pm\,4.6$ & $68.8\,\pm\,6.9^a$\\
\spitzer/MIPS & 70  & 4     & 18  & 10\%  & $1607.4\,\pm\,160.7$ & $1349\,\pm\,202^a$ \\
\spitzer/MIPS & 160 & 8     & 38  & 12\%  & $2601.9\,\pm\,317.1$ &- \\
\iras\ & 12  &  90  &  270  & 20\% & $58.8\,\pm\,12.6$ & $58\,\pm\,17^b$ \\
\iras\ & 25  &  90  &  276  & 20\% & $99.2\,\pm\,20.3$ & $90\,\pm\,27^b$  \\
\iras\ & 60  &  90  &  282  & 20\% &$1441.3\,\pm\,288.8$ & $1330\,\pm\,266^b$  \\
\iras\ & 100 &  90  &  300  & 20\% & $1759.2\,\pm\,356.2$ &
$1810\,\pm\,360^b$\\
\akari\  & 65  &  15   & 63  & 10\% & $1445.3\,\pm\,144.9$ &- \\
\akari\  & 90  &  15   & 78  & 10\% & $1735.5\,\pm\,173.7$ &- \\
\akari\  & 140 &  15   & 88  & 10\% & $2645.9\,\pm\,265.9$ &- \\
\akari\  & 160 &  15   & 88  & 10\% & $3114.1\,\pm\,313.6$ &- \\
\planck\ & 350  & 60  & 278 & 6.4\% & $623.7\,\pm\,50.5$ & $2003.5\pm144.2^c$\\
\planck\ & 550  & 60  & 290 & 6.1\% & $276.4\,\pm\,20.5$ & $728.8\pm51.4^c$\\
\planck\ & 850  & 60  & 296 & 0.78\% &$90.2\,\pm\,4.2$ & $230.3\pm16.2^c$\\
\planck\ & 1380  & 60  & 301 & 0.16\% & $32.3\,\pm\,1.1$& $56.2\pm4.0^c$\\
\hline
\end{tabular}
\tablerefs{\scriptsize (a) \citet{Koo2016}, (b) \citet{Mufson1986}, (c) \citet{Planck2016}.}
\end{table*}

\begin{table*}
\centering
\caption{The Dust Opacity and Estimated Dust Mass and Temperature from Infrared Emission}
\label{tab:kappa}
\begin{tabular}{c c c c c c c c}
\hline \hline
Literature & $\lambda_0^a$ & $\kappa_{\lambda_0}$ & $\beta$ & $M_{\rm w}$ &$T_{\rm w}$ & $M_{\rm c}$ & $T_{\rm c}$\\
  &  ($\rm \mu m$) &  ($\rm cm^2\,g^{-1}$) &   &  ($M_\odot$) & (K) & ($M_\odot$) & (K) \\
\hline
\cite{Hildebrand1983} & 125 & 18.9 & $1.53^{+0.22}_{-0.19}$ & $0.13^{+0.03}_{-0.03}$ & $55.1^{+2.8}_{-2.4}$ & $19.6^{+13.3}_{-7.2}$ & $20.9^{+2.2}_{-2.0}$ \\
\cite{James2002} & 850 & 0.7 & $1.53^{+0.21}_{-0.19}$ & $0.18^{+0.06}_{-0.05}$ & $55.1^{+2.8}_{-2.4}$ & $28.5^{+4.0}_{-3.3}$ & $20.8^{+2.2}_{-1.9}$ \\
\cite{Clark2016} & 500 & 0.51 &  $1.53^{+0.22}_{-0.19}$ & $0.56^{+0.16}_{-0.12}$ & $55.0^{+2.9}_{-2.4}$ & $88.6^{+22.7}_{-17.5}$ & $20.8^{+2.2}_{-2.0}$ \\
\hline
\cite{Draine1984}$^b$ & 500 & 1.45 & 2.04 & $0.10^{+0.06}_{-0.02}$ & $52.6^{+2.2}_{-5.7}$ & $45.9^{+10.3}_{-9.8}$ & $17.5^{+0.8}_{-0.7}$ \\

\cite{Li2001} & 500 & 0.95 &2.13 &$0.17^{+0.12}_{-0.03}$ &$50.0^{+2.4}_{-7.1}$  &$74.2^{+16.2}_{-15.8}$ & $17.1^{+0.8}_{-0.6}$  \\

\hline
\end{tabular}
\tablecomments{\scriptsize (a) The reference wavelength for normalization of mass absorption coefficient $\kappa_{\lambda_0}$. \scriptsize (b) The model dust opacity with $M_{\rm sil}/M_{\rm gra}=2:1$ used in Section \ref{sec:sed_fit}. }
\end{table*}

\end{CJK*}
\end{document}